# Effect of Fluorine doping on the electrocatalytic properties of $Nb_2O_5$ for $H_2O_2$ electrogeneration.

*Aline B. Trench [1], João Paulo C. Moura [1],*

*Caio Machado Fernandes [1], Mauro C. Santos [1*]*

[1] *Laboratório de Eletroquímica e Materiais Nanoestruturados, Centro de Ciências Naturais e Humanas, Universidade Federal do ABC. Rua Santa Adélia 166, Bairro Bangu, 09210-170, Santo André - SP, Brasil.*

*\*Corresponding author.* E-mail address: mauro.santos@ufabc.edu.br

# ABSTRACT

The oxygen reduction reaction (ORR) via the 2-electron mechanism is an efficient way to produce hydrogen peroxide ($H_2O_2$) under mild conditions. This study examines the modification of Vulcan XC72 carbon with fluorine (F)-doped niobium oxide ($Nb_2O_5$) nanoparticles at varying molar ratios (0, 0.005, 0.01, 0.02). The F-doped $Nb_2O_5$ nanoparticles were synthesized using the oxidizing peroxide method and then incorporated into Vulcan XC72 carbon via impregnation. Characterization techniques included X-ray diffraction (XRD), scanning electron microscopy (SEM), transmission electron microscopy (TEM), contact angle measurements, and X-ray photoelectron spectroscopy (XPS). Electrochemical evaluation using the rotating ring disk electrode method revealed that Vulcan XC72 modified with 1.0% F-doped $Nb_2O_5$ exhibited the best ORR performance. When used as a gas diffusion electrode, this electrocatalyst produced more $H_2O_2$ at all applied potentials than the pure and $Nb_2O_5$-modified Vulcan XC72 carbon. At potentials of -0.7 V and -1.3 V, the proposed electrocatalyst achieved $H_2O_2$ yields 65% and 98% higher than the $Nb_2O_5$-modified electrocatalyst. Furthermore, it presented lower energy consumption and higher current efficiency than the other electrocatalysts compared in this study. The enhanced performance is attributed to F doping, which increased $Nb_2O_5$ lattice distortion and disorder, improving electron availability for ORR. Additionally, F-doped electrocatalysts exhibited more oxygenated species and greater hydrophilicity, facilitating $O_2$ adsorption, transport, and electron transfer. These properties significantly enhanced $H_2O_2$ electrogeneration efficiency while reducing energy consumption.

# 1. Introduction

Hydrogen peroxide is a very important and environmentally friendly chemical compound. It is decomposed into water ($H_2O$) and oxygen ($O_2$). As a powerful oxidizing agent, it is widely used in various areas, such as health, industry, and the environment. It is used as a disinfectant, antiseptic, and bleaching agent, and in the treatment of effluents [1].

Currently, the industrial production of hydrogen peroxide occurs mainly through anthraquinone. This process involves the steps of (1) hydrogenation of anthraquinone to form the corresponding hydroquinone, (2) oxidation of hydroquinone with oxygen from the air, forming $H_2O_2$, and (3) extraction and purification, where the $H_2O_2$ is separated and concentrated. Despite being broadly used, this method of producing $H_2O_2$ has disadvantages such as excessive energy consumption due to the multiple steps involved, high temperatures and pressures, which make the process more expensive, and the use of toxic and flammable solvents, which pose safety risks [2, 3]. Many studies are looking for more sustainable methods to reduce costs and environmental impact in this scenario. The oxygen reduction reaction (ORR) is a strategic solution to the anthraquinone reaction [4, 5]. It can follow 2- and 4-electron mechanisms, with the 2-electron mechanism in an acidic medium leading to the formation of hydrogen peroxide, as shown in equation (1), and the 4-electron mechanism in an acidic medium driving to the formation of $H_2O$, as shown in equation (2) [6-8]:

$$O_2 + 2e^- + 2H^+ \rightarrow H_2O_2 \qquad (1)$$

$$O_2 + 4H^+ + 4e^- \rightarrow 2H_2O \qquad (2)$$

ORR is more advantageous than the traditional method of anthraquinone, being a sustainable, environmentally friendly process carried out in situ, without generating hazardous waste. In addition, it operates at low pressure and temperature, which reduces energy costs. Among the catalysts used in ORR, gold (Au) stands out for its efficiency in generating $H_2O_2$. However, its high cost and limited availability make its use unviable [9].

Carbon materials are an alternative catalyst to noble metal catalysts, such as Au, in the two-electron ORR to produce hydrogen peroxide. Carbonate materials have good selectivity for hydrogen peroxide, low cost, electrochemical stability, high surface area, and porosity [10-15].

In addition, they can be easily modified with metal oxides, further improving their selectivity for $H_2O_2$ generation. Metal oxides such as $WO_3$ [16], $MnO_2$ [17, 18], $CeO_2$ [19], $ZrO_2$ [20], $V_2O_5$ [21], among others, have already been reported. Their inclusion increases the hydrophilicity and number of oxygenated species, facilitating the transport and adsorption of $O_2$, which favors the 2-electron mechanism for the electrogeneration of $H_2O_2$.

Niobium (Nb) exists as stoichiometric oxides such as NbO, $NbO_2$, and $Nb_2O_5$, the latter being the most explored [22]. $Nb_2O_5$ is an n-type semiconductor with a band gap of 3.4 eV. It has a wide range of applications such as gas sensors, catalysts, and solar cells. [23, 24]. Carneiro *et al.* [25] used reduced graphene oxide sheets modified with $Nb_2O_5$ in a "bean" shape for $H_2O_2$ generation via ORR, where the authors found that the modified material exhibited higher electrocatalytic activity for ORR. Trench *et al.* [26] reported that Vulcan XC72 carbon modified with $Nb_2O_5$ produced about 48 % more $H_2O_2$ than unmodified carbon. This response may be associated with the greater hydrophilicity presented by the catalysts modified with $Nb_2O_5$ and a more significant number of oxygenated species.

One strategy that can improve the properties of a semiconductor is doping. Doping occurs when intentionally added atoms or ions replace the original anions or cations in the structure of a material, causing changes in its electronic, structural, and chemical properties [27-30].

In a recent study, doping $Nb_2O_5$ with Ce proved to be an efficient strategy for improving the catalytic efficiency of Vulcan XC 72 carbon-based electrocatalysts. Vucaln XC 72 carbon electrocatalysts modified with pure $Nb_2O_5$ and with Ce-doped $Nb_2O_5$ were compared. The electrocatalysts containing Ce-doped $Nb_2O_5$ achieved an electrogeneration of $H_2O_2$ of approximately 105% that of the catalysts with $Nb_2O_5$. In addition, the doped electrocatalysts presented a lower energy cost and a higher current efficiency [10].

In this study, Vulcan XC72 carbon modified with $Nb_2O_5$ doped with F in different molar amounts is presented as a novel electrocatalyst for $H_2O_2$ electrogeneration from ORR. The synthesized materials will be characterized by XRD, SEM, TEM, contact angle, and XPS, and evaluated for their performance for ORR. A GDE of the best ORR catalyst will be fabricated to analyze the electrogeneration of $H_2O_2$. The electrogenerated quantity of $H_2O_2$ at different potentials will be evaluated, together with the energy costs and current efficiency. This study explores how F doping in the $Nb_2O_5$ crystal lattice can affect the efficiency of ORR catalysts for $H_2O_2$ electrogeneration, carefully assesses the results observed in the characterization techniques, and relates them to the results applied to $H_2O_2$ electrogeneration.

## 2. Materials and methods

2.1. Preparation of the electrocatalysts

The fluorine (F)-doped $Nb_2O_5$ nanorods were synthesized using the peroxide oxidation method [31]. Initially, the compound $(NH_4[NbO(C_2O_4)_2(H_2O)]\cdot(H_2O)_n$ (Sigma Aldrich) was dissolved in 20 mL of distilled water under intense magnetic stirring. Then, a specific amount of ammonium fluoride (Sigma Aldrich) was added to this solution, corresponding to 0, 0.005, 0.01, and 0.02 moles, and the mixture was kept under stirring for 10 minutes. Subsequently, hydrogen peroxide ($H_2O_2$, Sigma Aldrich) was placed into the solution in the molar ratio of $H_2O_2$: $NH_4[NbO(C_2O_4)_2(H_2O)]\cdot(H_2O)_n$ of 10:1, followed by another 10 minutes of vigorous stirring. Since high dopant concentrations can lead to a large change in the crystal structure of the material and lead to the formation of secondary phases, the maximum dopant amount chosen in this study was 0.02 moles [32]. Furthermore, $Nb_2O_5$ doped at low concentrations proved to be more efficient in the electrogeneration of $H_2O_2$ compared to higher doping levels [10].

The solution obtained was then added to a hydrothermal reactor and subjected to heat treatment at 140°C for 12 h. After cooling, the material was centrifuged, washed with distilled water, and dried at 80°C for 12 h. The materials received their names according to the amount of dopant, being: pure $Nb_2O_5$, Nb/F 0.5%, Nb/F 1.0%, and Nb/F 2.0%, corresponding to concentrations of 0.000, 0.005, 0.01, and 0.02 moles, respectively.

To prepare the electrocatalysts, the F-doped $Nb_2O_5$ nanoparticles were placed on a Vulcan XC-72 carbon support using the impregnation method [33]. For this purpose, 0.5 g of Vulcan XC-72 carbon was added to 30 mL of water under magnetic stirring for 15 min. The $Nb_2O_5$ and F-doped $Nb_2O_5$ nanoparticles were weighed to produce electrocatalysts containing 1% by mass of these nanoparticles, which were then added to the Vulcan XC-72 carbon dispersions and kept under magnetic stirring for 4 h. Subsequently, the electrocatalysts were dried in an oven at 90°C.

The electrocatalysts obtained were named VC/Nb, VC/Nb/F 0.5%, VC/Nb/F 1.0%, and VC/Nb/F 2.0%, corresponding to doping concentrations of 0.000, 0.005, 0.01, and 0.02 moles, respectively.

2.2. Characterization of electrocatalysts

X-ray diffraction (XRD) analyses were performed using a D8 Focus diffractometer manufactured by Bruker AXS. The X-ray source used was a copper Kα (wavelength λ = 1.54 Å) operating in continuous scanning mode at a rate of 2 degrees per minute, covering the range of 20 to 80 degrees (2θ).

To obtain the band gap energy of the materials, a UV-Vis spectrophotometer (VARIAN 50 Scan) was used in diffuse reflectance mode.

Morphological analysis was conducted by scanning electron microscopy (SEM) using a JEOL FESEM microscope. Transmission electron microscopy (TEM) was performed using a JEOL JEM 2100 microscope. Energy dispersive X-ray spectroscopy (EDS) generated elemental maps in scanning transmission electron microscopy (STEM) mode.

Contact angle measurements were performed using a goniometer (GBX Digidrop). For this measure, 2 mg/mL of the material was prepared in distilled water and sonicated for 60 seconds in an ultrasonic bath. Then, a 20 µL aliquot of each dispersion was deposited on a glassy carbon plate and dried to form a uniform film. Subsequently, 20 µL of distilled water was gently applied to the film's surface to determine the contact angle. Measurements were performed using Windrop software.

X-ray photoelectron spectroscopy (XPS) was performed on a Scienta Omicron ESCA+ spectrometer (Germany), using monochromatic Al Kα X-ray radiation (energy of 1486.7 eV) as the source. The Shirley method subtracted the inelastic background in the high-resolution spectra of the C 1s level. Spectral fitting was performed without restrictions using multiple Voigt profiles, and data analysis was conducted using CasaXPS software.

### 2.3. Oxygen reduction reaction study

For the electrochemical measurements, an Autolab PGSTAT-302N potentiostat/galvanostat, operated with NOVA software, in a rotating ring disk electrode (RRDE) system, was used. The ORR was tested using a 125 mL cell containing a platinum (Pt) counter electrode with a surface area of 2 cm², a Hg/HgO reference electrode containing an internal solution of NaOH 5 mol L$^{-1}$, and the RRDE itself as the working electrode. In an alkaline medium, the hydroperoxyl ion radical is produced in the 2-electron mechanism, being the most stable system, presenting less noisy currents; this medium is chosen for ORR tests [19]. The RRDE is composed of a glassy carbon disk with an area of 0.2475 cm² and a platinum (Pt) ring with an area of 0.1866 cm², with a collection factor N = 0.26. The experiments were performed in an electrolyte solution of 100 mL of NaOH 1 mol L$^{-1}$ (Synth). The $Nb_2O_5$-modified VC XC72 carbon-based electrocatalysts with and without fluorine doping were applied to the RRDE disk as dispersions prepared at a ratio of 1 mg mL$^{-1}$ (material/distilled water) using high-power ultrasound for 1 min. 20 µL of each dispersion was then deposited on the glassy carbon substrate, and the water was evaporated. After this step, 20 µL of Nafion solution (1:100 v/v, Nafion: distilled water, Sigma Aldrich) was applied to the dried film and dried again.

For all electrochemical analyses, the electrolyte was saturated with oxygen gas ($O_2$, GasNorte) for 30 min, maintaining the $O_2$ flow over the solution during the measurements. The experiments were conducted at a scan rate of 5 mV s$^{-1}$ at room temperature. The ring electrode was polarized at 0.3 V (vs Hg/HgO) to ensure total oxidation of the $HO_2^-$ ions that reached it, while the disk electrode was subjected to an anodic scan between 0.1 V and -0.6 V (vs Hg/HgO). The rotations were controlled by a CTV10 speed control unit [18, 34].

2.4. $H_2O_2$ electrogeneration

Using the electrocatalyst with the best response to ORR, a gas diffusion electrode (GDE) measuring 2 cm in diameter and 0.4 cm thick was constructed, providing an exposed surface area of 3 cm² between two porous steel plates. For its manufacture, 6 grams of the electrocatalyst were dispersed in 100 mL of $H_2O$ for 30 min. Then, a polytetrafluoroethylene (PTFE) solution containing 60% PTFE by mass (Sigma Aldrich) was added and stirred with 20 mL of distilled water in a ratio of 4:1 (material: PTFE). This solution was continuously stirred for 40 min. The mixture obtained was subjected to vacuum filtration and dried until reaching a moisture content between 5% and 10%. Subsequently, the material was compacted between two steel plates at 290°C for 2 h, using a thermo-hydraulic press (SOLAB SL-10/15-E). To control the potential applied during the electrogeneration of $H_2O_2$, an Autolab PGSTAT-302N potentiostat/galvanostat was used. In addition, a multimeter was used to measure the cell potential.

The electrochemical experiments were conducted in a cell containing 350 mL of a supporting electrolyte solution composed of 0.1 mol L⁻¹ of $H_2SO_4$ (Synth) and 0.1 mol L⁻¹ of $K_2SO_4$ (Synth). The cell was equipped with a coil connected to a thermostatic bath to maintain a constant temperature of 20°C. The system was configured with a platinum auxiliary electrode (anode) with an area of 7.5 cm², an Ag/AgCl (saturated KCl) reference electrode, GDE acting as the working electrode (cathode), and the distance between the cathode and the anode was fixed at 1 cm.

In $H_2O_2$ electrogeneration tests by the GDE, different potentials were applied for 2 hours. Before starting the application of potential, pressurized $O_2$ (0.2 bar) was purged directly into the GDE for 30 min to ensure saturation of the electrolyte with $O_2$ during the analysis. The evolution of $H_2O_2$ electrogeneration was accompanied by collecting 0.5 mL aliquots of the solution. These samples were added to 4 mL of an ammonium molybdate (($NH_4$)$_6Mo_7O_{24}$) solution at 2.4 mmol L$^{-1}$ in $H_2SO_4$ 0.5 mol L$^{-1}$. The analyses were performed in a UV-Vis spectrophotometer (VARIAN 50 Scan), measuring the absorbance at 350 nm to calculate the $H_2O_2$ concentration. The quantification was expressed in mg L$^{-1}$, with a correlation coefficient of 0.99.

## 3. Results and discussion

Fig. 1(a) shows the XRD patterns of pure and F-doped $Nb_2O_5$ at molar percentages of 0.0, 0.5, 1.0, and 2.0%. Diffraction peaks corresponding to the orthorhombic phase of $Nb_2O_5$ (JCPDS 28-0317) are observed. The peak named θ at 2θ ≈ 28° is related to the presence of hydrated niobium oxide due to intermediate crystallization conditions [35]. The XRD patterns for the doped samples did not change and remained like those of pure $Nb_2O_5$. This behavior indicates that F doping did not generate secondary phases, maintaining the crystalline structure of $Nb_2O_5$.

Figure 1(b) shows that the diffraction peaks of the F-doped samples are shifted to a lower angle, indicating that the crystal lattice of the doped samples may have been expanded by doping. This expansion can be explained by the compensation charged for F-doping. The $Nb^{5+}$ ions may have been reduced to $Nb^{4+}$ ions to compensate for the insertion of the F$^-$ ion. Thus, since the ionic radius of $Nb^{4+}$ (0.68 Å) is larger than that of Nb$^{5+}$ (0.64 Å), the crystal lattice of F-doped may have been enlarged. Furthermore, it can be noted that the Nb/F 2.0% sample presents a smaller deviation for lower angles than the other samples. This can be explained by the saturation of the F incorporation capacity in the Nb$_2$O$_5$ crystal structure, which may have led to the formation of vacancies, causing a small contraction of the lattice compared to other doped samples [36-38].

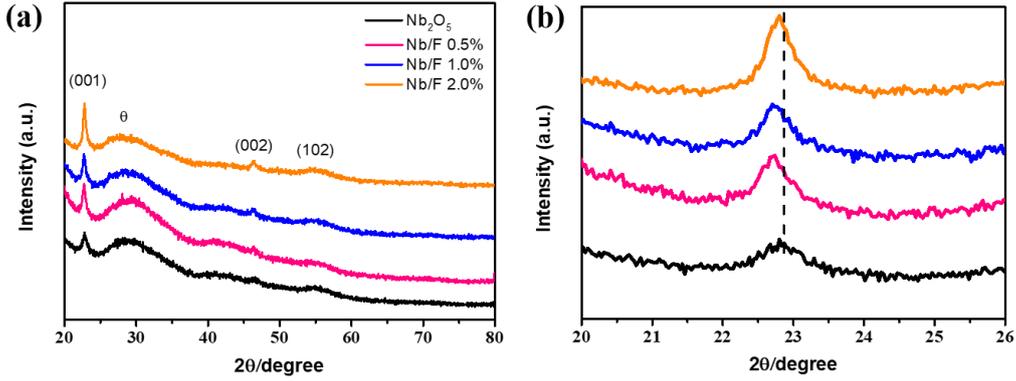

**Fig. 1.** XRD patterns of Nb$_2$O$_5$ and Nb/F with different amounts of F (a) and amplified region of the XRD patterns (b).

The band gap energy (E$_{gap}$) of the Nb$_2$O$_5$ and F-doped Nb$_2$O$_5$ samples was calculated using the Tauc method, following Equation (3) [39]:

$$(\alpha h\nu)^{2/n} \sim h\nu - E_g, \qquad (3)$$

where n = 1 for a semiconductor with direct E$_{gap}$ and n = 4 for a semiconductor with indirect E$_{gap}$, α = absorbance, and *hv* is the photon energy. As can be seen in Fig. S1, Nb$_2$O$_5$ presented a direct E$_{gap}$ of 3.46 eV, a value close to that reported in the literature [40]. The samples doped with 0.5, 1.0, and 2.0% F presented an E$_{gap}$ of 3.53, 3.65, and 3.66 eV, respectively, demonstrating that with the increase in the amount of dopant, there was an increase in the E$_{gap}$ value. This behavior can be attributed to the Burstein-Moss effect [41]. In this effect, doped n-type semiconductors cause the Fermi level to move upwards in the conduction band, filling the lower part of the semiconductor conduction band with electrons [42].

SEM micrographs of $Nb_2O_5$ and F-doped $Nb_2O_5$ samples are presented in Fig. 2. All samples present a morphology defined by non-uniform agglomerated particles [26], showing that F doping did not have an evident influence on the morphology of the samples. This behavior was also reported for Ce-doped $Nb_2O_5$ samples [10].

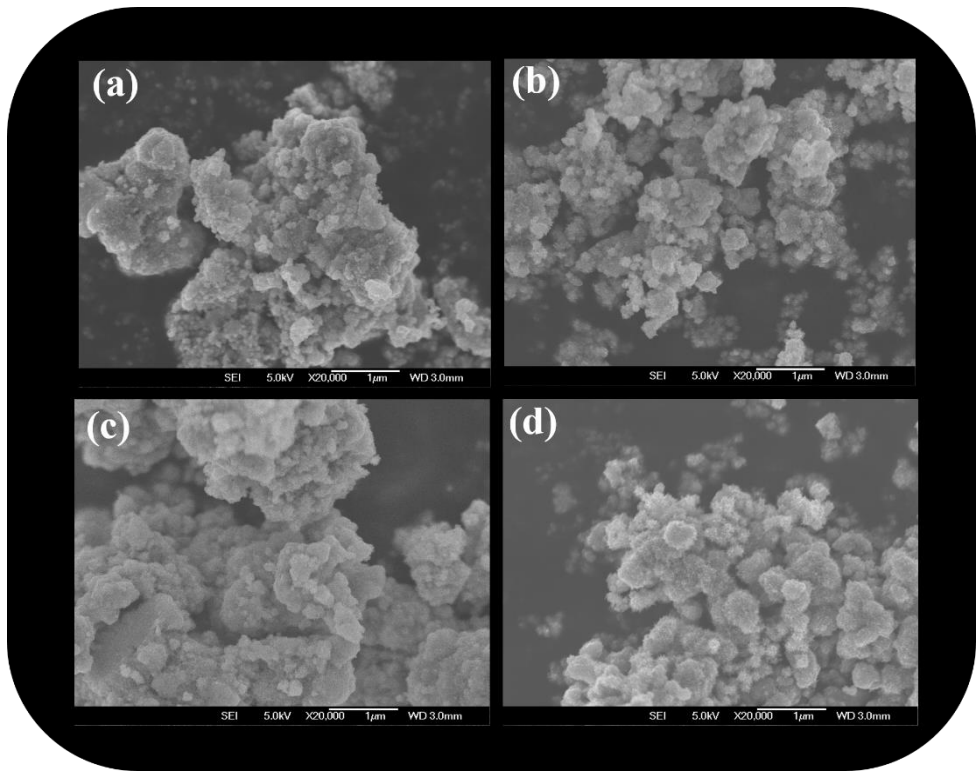

**Fig. 2.** SEM images of (a) $Nb_2O_5$, (b) Nb/F 0.5%, (c) Nb/F 1.0 %, and (d) Nb/F 2.0%.

Fig. 3 shows TEM images obtained for the $Nb_2O_5$ sample. It can be seen in Fig. 3(a) a cluster of nanoparticles, which when analyzed at high resolution (HRTEM) (Fig. 3(b)), has the shape of nanorods. Fig. 3(c) presents the corresponding Fast Fourier Transform (FFT) of the highlighted area in red lines shown in Fig. 3(b). The FFT analysis indicated the presence of $Nb_2O_5$, based on the (001) plane with an interplanar distance of 3.96 Å [43, 44]. This result is in agreement with those observed in the XRD analyses, reinforcing that $Nb_2O_5$ was successfully synthesized.

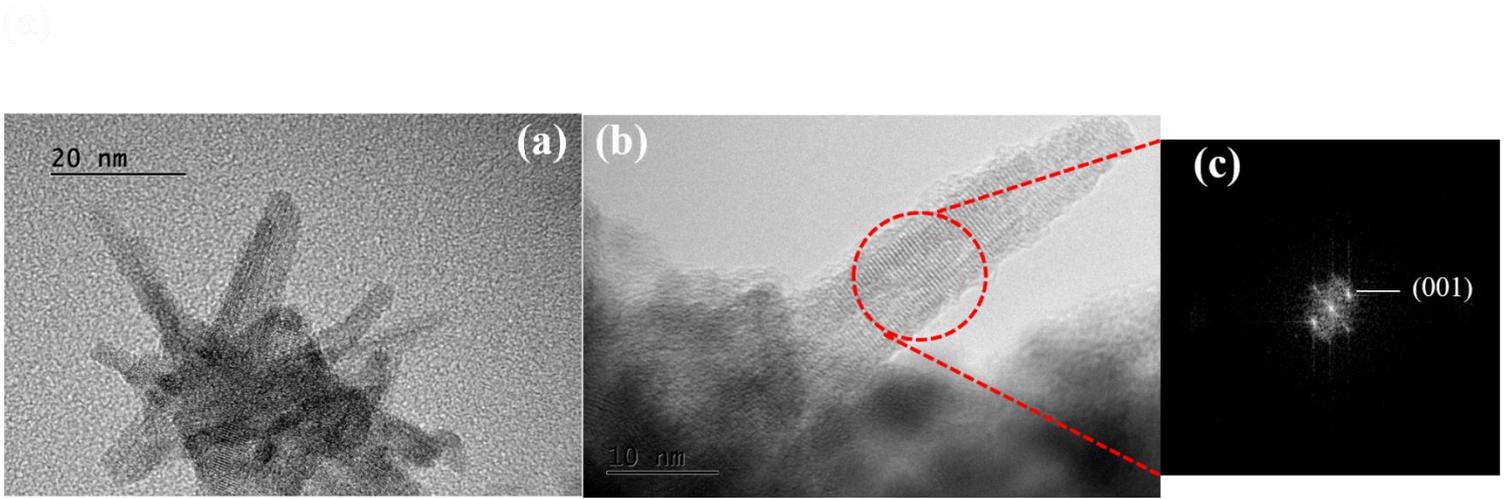

**Figure 3.** TEM characterization of $Nb_2O_5$ under (a) low magnification, (b) HRTEM image, and (c) its corresponding FFT.

Fig. 4 presents TEM images of the $Nb_2O_5$ sample with a higher percentage of dopant (2% F). In Fig. 4(a), we can notice many nanorod-shaped particles, similar to those observed for the undoped $Nb_2O_5$ sample. In Fig. 4(b) of HRTEM, the nanorods are even more evident. The corresponding FFT of the highlighted area in red lines in Fig. 4(b) is shown in Fig. 4(c). The FFT analysis showed an interplanar distance of 3.84 Å, which also corresponds to the (001) plane of $Nb_2O_5$. This demonstrates efficient doping of F in the $Nb_2O_5$ crystal structure.

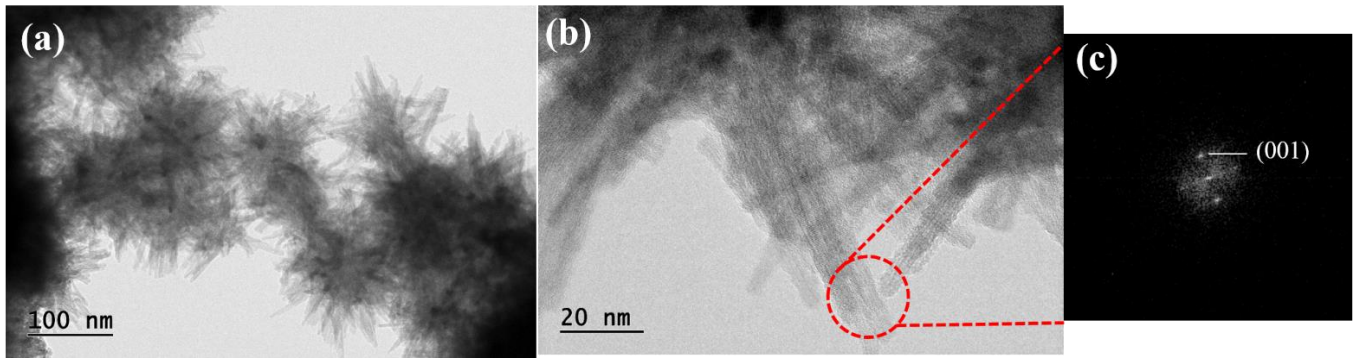

**Figure 4.** TEM characterization of Nb/F 2.0% under (a) low magnification, (b) HRTEM image, and (c) its corresponding FFT.

To confirm the presence of F in the $Nb_2O_5$ crystal structure, an analysis of the elemental composition of the Nb/F2.0% sample was performed by EDS mapping. Fig. 5 presents the results of this mapping, where we can see that the nanorod clusters are composed of Nb, O, and F, uniformly distributed. The presence of F confirms the insertion of F in the $Nb_2O_5$ crystal structure in the form of a dopant.

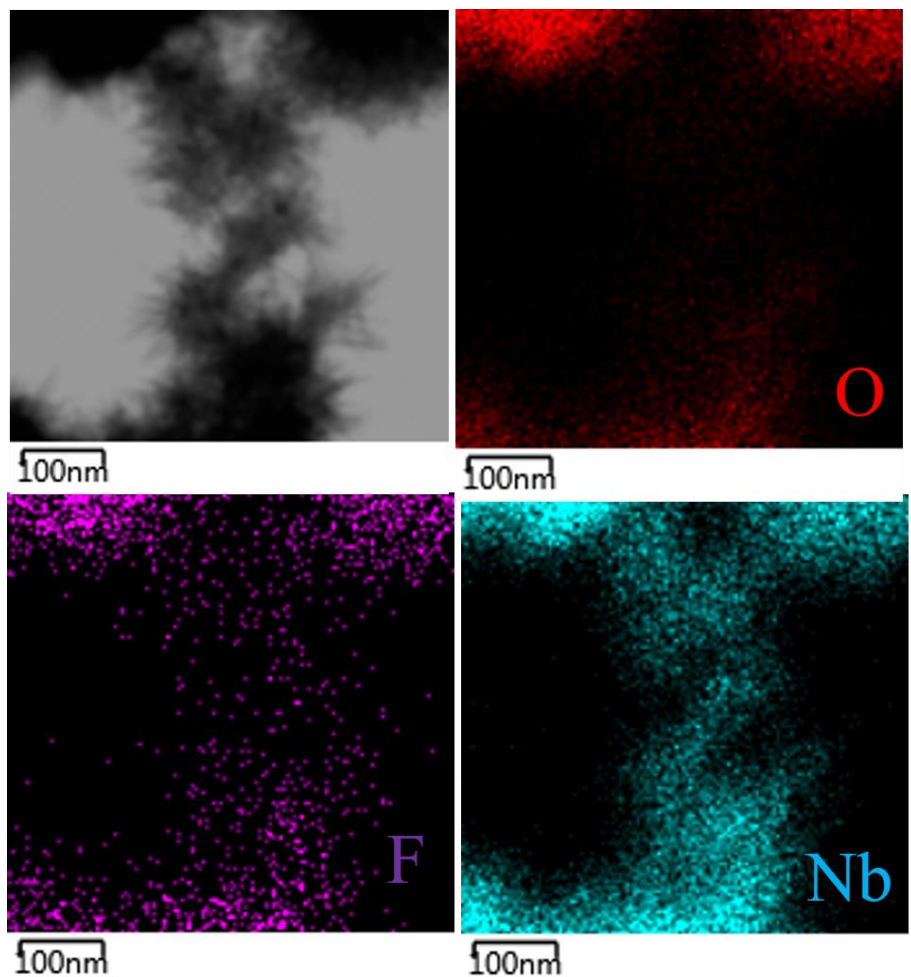

**Fig. 5.** EDS elemental mapping of O, Nb, and F for the sample Nb/F 2.0%.

The synthesized samples of $Nb_2O_5$ and $Nb_2O_5$ doped with F in different molar percentages were impregnated in VC XC72, and these catalysts were analyzed for their hydrophilicity by contact angle measurements. The average contact angle values found are listed in Table 1. It can be observed that the V/Nb catalyst presented a lower contact angle value than the VC XC72 catalyst, indicating that the VC/Nb catalyst presents a greater hydrophilicity. When the VC/Nb/F catalysts are analyzed, they present lower contact angle values than VC/Nb. Thus, the $Nb_2O_5$ catalysts doped with F presented a greater hydrophilicity compared to the pure $Nb_2O_5$ catalysts. This greater hydrophilicity may be associated with a greater number of oxygenated carbon groups, which results in greater wettability, which may favor ORR [18, 34].

**Table 1.** Average contact angles of the synthesized electrocatalysts.

| Electrocatalyst | Contact Angle (°) | |
|---|---|---|
| VC XC72 | 86.66 | 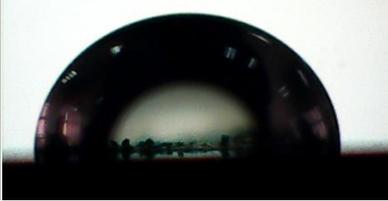 |
| VC/Nb | 68.48 | 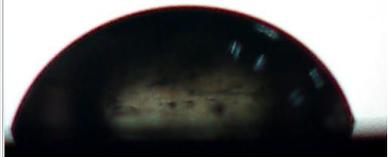 |
| VC/Nb/F 0.5% | 66.39 | 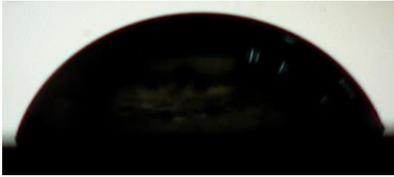 |
| VC/Nb/F 1.0% | 60.71 | 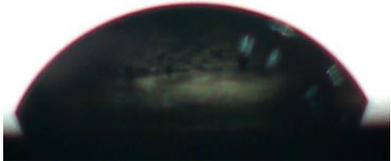 |
| VC/Nb/F 2.0% | 67.98 | 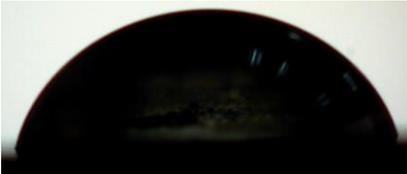 |

XPS analysis is performed to analyze the chemical elements present in the catalysts. In the Nb 3d spectra of VC/Nb, VC/Nb 0.5%, VC/Nb 1.0%, and VC/Nb 2.0%, two distinct peaks are observed at approximately 207 and 210 eV, which are attributed to Nb $3d_{5/2}$ and Nb $3_{d3/2}$, respectively, of the $Nb^{5+}$ oxidation state [45]. In the F-doped catalysts, the formation of a new component related to $Nb^{4+}$ state can be observed, which is in agreement with the results observed in the XRD analyses. In addition, the spectra of all F-doped catalysts undergo a shift to higher energies of approximately 0.4 eV, also related to the emergence of the Nb$^{4+}$ component. These results may suggest that Nb was reduced during doping, introducing electron donor states, resulting in n-type doping [46], generating greater availability of free electrons and catalyst reactivity, which may favor the electron transfer necessary for the ORR.

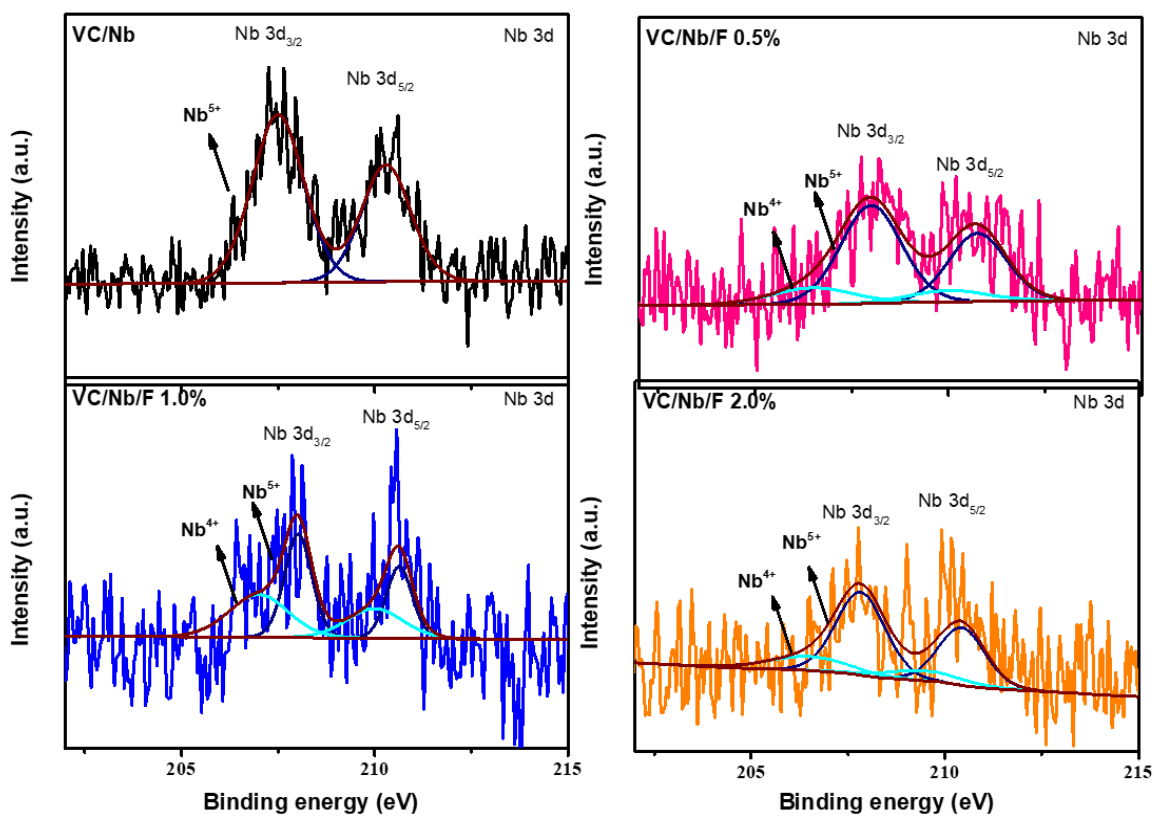

**Fig. 6.** Deconvoluted Nb 3d XPS spectra for VC/Nb, VC/Nb/F 0.5%, VC/Nb/F 1.0%, and VC/Nb/F 2.0%.

Fig. 4 shows high-resolution XPS spectra of C 1s of the VC/Nb, VC/Nb/F 0.5%, VC/Nb/F 1.0%, and VC/Nb/F 2.0%. The spectra were deconvoluted into four components, with the component located at approximately 284 eV corresponding to the C-C bond, and the components at approximately 285, 286, and 289 eV related to the C–O, C=O, and COOH bonds, respectively [47, 48]. It can be noted that the Vulcan XC72 catalyst presented 54.60 at.% of oxygenated species, while the VC/Nb catalyst presented 62.88 at.%. When we analyze the catalysts doped with F, an increase in oxygenated species can be noted, especially for the VC/Nb/F 1.0% catalyst, which presented 69.41 at.%. These results indicate that modification of Vulcan XC72 carbon materials with F-doped $Nb_2O_5$ nanoparticles can provide an increase in oxygenated species on the carbon surface, making it more hydrophilic. These results are in good agreement with those observed previously from contact angle measurements.

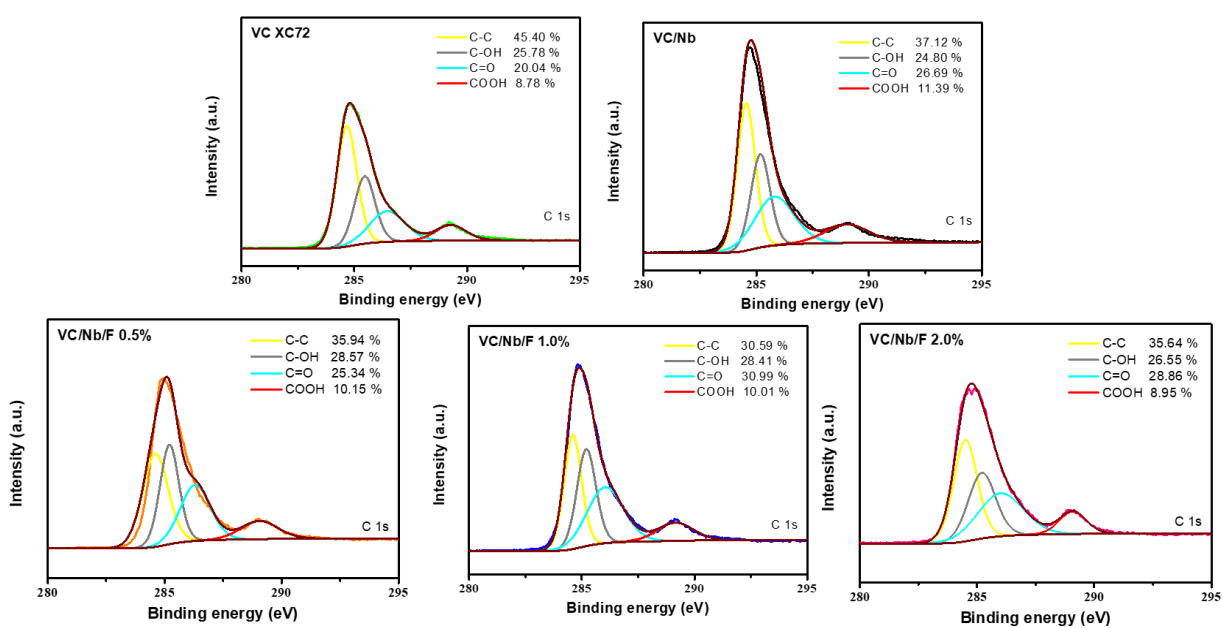

**Fig. 7.** Deconvoluted C 1s XPS spectra of VC/Nb, VC/Nb/F 0.5%, VC/Nb/F 1.0%, and VC/Nb/F 2.0%.

Fig. S2 shows the deconvoluted O 1s spectra for the synthesized electrocatalysts. The O 1s spectrum presented components related to the C=O (532 eV), O-C=O (533 eV), and water (534 eV) groups. In addition, for the electrocatalysts modified with $Nb_2O_5$, a component related to the O-Nb bonds (530 eV) was also observed [10]. VC XC72, VC/Nb, and VC/Nb/F 1.0% electrocatalysts presented C=O and O-C=O group concentrations of 86.83, 91.40, and 93.25.34 at.%, respectively. These results are similar to those observed in the C 1s spectrum, highlighting the greater number of oxygenated species in the sample doped with 1% F.

The electrocatalysts produced in this study, as well as the reference electrocatalysts of VC XC72 and Pt/C, were evaluated for ORR in alkaline medium. The results obtained are shown in Fig. 8 and were performed under RRDE rotation set to 1600 rpm. The VC/Nb and VC/Nb/F electrocatalysts exhibited disk currents close to those of VC XC72 and lower than those of Pt/C. This allows us to conclude that the developed electrocatalysts follow the 2-electron mechanism, since VC XC72 is the reference material for the 2-electron ORR mechanism. Furthermore, all VC/Nb/F catalysts showed higher currents than those obtained for VC/Nb, VC XC72 and Pt, indicating that these catalysts are efficient for $H_2O_2$ electrogeneration. In particular, the VC/Nb/F 1.0% electrocatalyst showed a current of 60 µA, while VC/Nb only 45 µA, indicating that F doping favored the 2-electron mechanism for RRO.

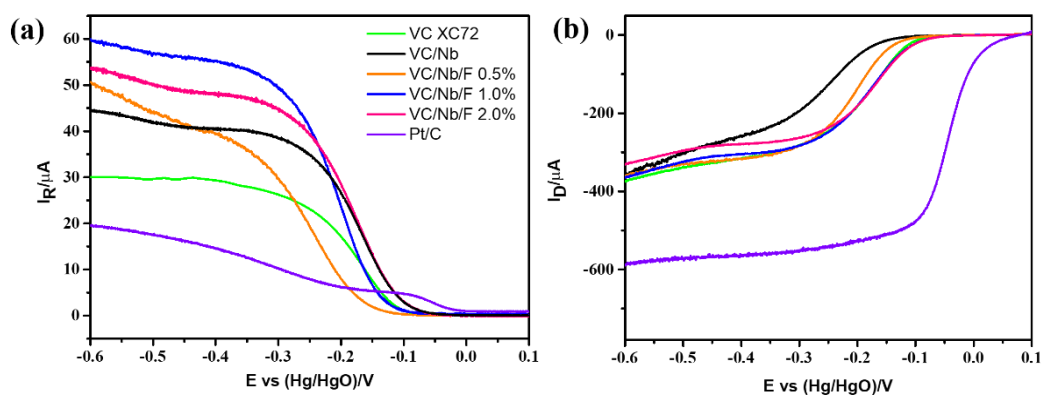

**Fig. 8.** Steady-state polarization curves for ORR electro-catalyzed by electrocatalysts in 1 mol L$^{-1}$ NaOH saturated with O$_2$ at a sweep rate of 5 mV s$^{-1}$ using a RRDE at 1600 rpm: (a) Ring current (Ering = 0.3 V ) and (b) Disc current.

The percentages of $H_2O_2$, $H_2O$ production, and the number of transferred electrons ($n_t$) were analyzed by the polarization curves obtained at 1600 rpm and applied in Eqs. 4 to 6 below, in which ir represents the ring current, id represents the disk current and N represents the current collection efficiency of the gold ring (with N equal to 0.26) [49] .

$$H_2O_2\% = \frac{200\ ir/N}{id + ir/N} \tag{4}$$

$$H_2O\ \% = 100 - H_2O_2\% \tag{5}$$

$$n_t = \frac{4id}{id + ir/N} \tag{6}$$

The calculation results for the electrocatalysts are presented in Table 2. The VC XC72 and Pt/C electrocatalysts presented values of number of transferred electrons ($n_t$) of 2.9 and 3.65, values close to those expected for these reference materials for the 2 and 4 electron mechanisms, respectively. The VC/Nb and VC/Nb/F electrocatalysts presented results closer to 2 than the reference material, demonstrating that these synthesized electrocatalysts are more selective than VC XC72 for the 2 electron mechanism for ORR. Table 2 also shows the percentages of $H_2O_2$ produced for each electrocatalyst. It can be noted that all the synthesized electrocatalysts present higher $H_2O_2$ production than VC XC72. However, the F-doped $Nb_2O_5$ electrocatalysts were more efficient than the electrocatalysts containing only $Nb_2O_5$, with the electrocatalyst containing 1% F (VC/Nb/F 1.0%) being the most efficient, producing 80% of $H_2O_2$, while the electrocatalyst containing only $Nb_2O_5$ (VC/Nb) produced 70%. Analyzing the previous results, it is noted that the VC/Nb/F 1.0% electrocatalyst presented a lower contact angle value and a higher concentration of oxygenated species in the XPS analyses. These results reflect the greater efficiency of this electrocatalyst for ORR, since electrocatalysts with greater amounts of oxygenated species and more hydrophilic have greater ease of transport and adsorption of $O_2$, facilitating the electrogeneration of $H_2O_2$ [20, 34, 49].

Furthermore, F doping may have caused an expansion of the $Nb_2O_5$ crystal lattice, as observed in XRD and XPS, generating local distortions, in addition to a greater number of free electrons available to act in the ORR [10, 36]. It is important to highlight that F doping above 1% had its efficiency reduced, indicating that the value of 1% is ideal for ORR applications.

**Table 2.** Number of electrons transferred ($n_t$), $H_2O$ %, and $H_2O_2$ % in ORR.

| Electrocatalyst | $n_t$ | $H_2O$ (%) | $H_2O_2$ (%) |
|---|---|---|---|
| VC XC72 | 2.9 | 47 | 53 |
| VC/Nb | 2.6 | 29.5 | 70.5 |
| VC/Nb/F 0.5% | 2.55 | 27.5 | 72.5 |
| VC/Nb/F 1.0% | 2.4 | 20.0 | 80.0 |
| VC/Nb/F 2.0% | 2.5 | 25.4 | 74.6 |
| Pt | 3.65 | 83.8 | 16.2 |

With the results observed above, the study continued to produce a GDE with the electrocatalyst with the best performance, choosing the VC/Nb/F 1.0% electrocatalyst, named VC/Nb/F1GDE. The results of the electrogeneration of $H_2O_2$ (mg L$^{-1}$) at various applied potentials and using the produced GDE are presented in Fig. 9. It can be noted that at the applied potential of -2.5 V, the VC/Nb/F1GDE produced 907.61 mg L$^{-1}$ of $H_2O_2$ in 2 h of electrolysis. Trench et al.[26] reported the use of Vulcan carbon GDE modified only with $Nb_2O_5$, named Nb1GDE, under the same experimental conditions for electrogeneration of $H_2O_2$. The results of this study are compared with the present study in Table 3. It can be noted that at all applied potentials, the VC/Nb/F1GDE produced more $H_2O_2$ than the Nb1GDE, demonstrating that F doping favors the ORR. In particular, at the potential of -2.5 V, the VC/Nb/F1GDE produced about 40% more $H_2O_2$ than the $Nb_2O_5$ GDE without the F dopant.

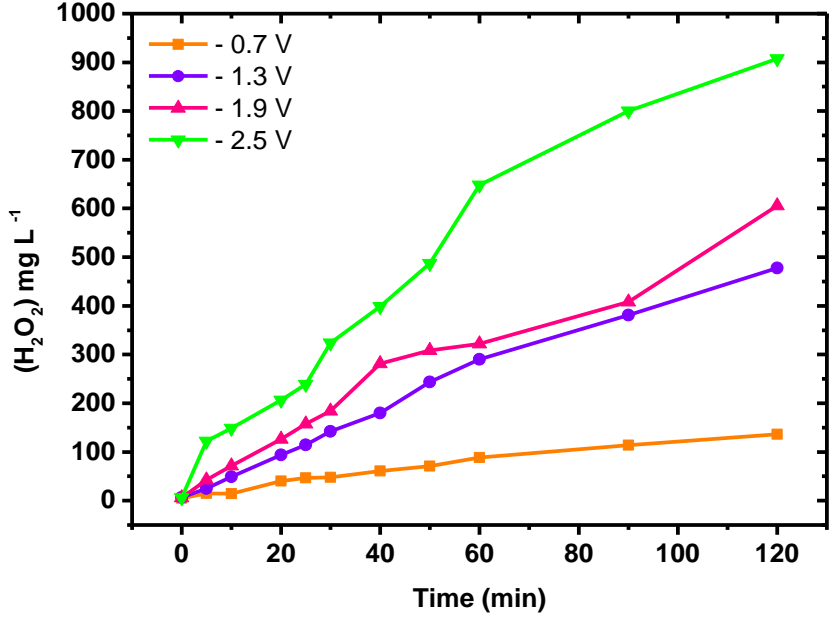

**Fig. 9.** $H_2O_2$ concentration obtained using VC/Nb/F1GDE at different potentials in 0.1 mol L$^{-1}$ $H_2SO_4$ and 0.1 mol L$^{-1}$ $K_2SO_4$ after 120 min of electrolysis.

Eqs. 7 and 8 below were used to calculate the energy consumption (EC, in kWhkg$^{-1}$) and the current efficiency (EC, in %). For Eq. 7, I is the current measured during electrolysis (in A), Ecel is the cell potential (in V), t is the electrolysis time (in h) and m is the mass of electrogenerated H2O2 [50]. In Eq. 8, z is considered the number of electrons in the ORR intended for the electrogeneration of $H_2O_2$ (z = 2), F is the Faraday constant, C $H_2O_2$ is the concentration of electrogenerated $H_2O_2$ (in g L$^{-1}$), V is the volume of the solution (in L), M $H_2O_2$ is the molar mass of $H_2O_2$ (34.01 g mol$^{-1}$), I is the current measured during electrolysis (in A) and t is the electrolysis time (in s) [50].

$$EC = \frac{I\,E_{cel}\,t}{1000\,m} \quad (7)$$

$$CE = \frac{z F C_{H2O2}\,V}{M_{H2O2}\,It}\,100 \quad (8)$$

Table 3 also presents the results obtained for EC and EC VC/Nb/F1GDE and Nb1GDE [26] after 120 min of electrolysis at different applied potentials. It can be noted that for all applied potentials, VC/Nb/F1GDE presented lower EC and higher CE values than Nb1GDE, demonstrating that the developed electrocatalyst is more selective for 2-electron ORR, producing large amounts of $H_2O_2$ at a lower energy cost and greater current efficiency. The great catalytic response of this catalyst is related to the insertion of F as a dopant in the $Nb_2O_5$ crystal lattice, which caused an increase in the $Nb_2O_5$ crystal lattice and local distortions, increasing the degree of disorder of this electrocatalyst, in addition to having increased the number of free electrons to act in the ORR. In addition, the electrocatalysts doped with F presented a greater number of oxygenated species and a more hydrophilic character, which are intrinsically related to greater ease of electron transfer, transport, and adsorption of $O_2$, enhancing the electrogeneration of $H_2O_2$ [18, 19, 51].

**Table 3.** Energy consumption (EC) and current efficiency (CE) obtained for VC/Nb/F1.0GDE based on their application at different potentials.

| GDE | Applied potential (vs. Ag/AgCl) | $H_2O_2$ (mg L$^{-1}$) | EC (kWh kg$^{-1}$) | CE (%) | Reference |
|---|---|---|---|---|---|
| Nb1GDE | -2.5 V | 645.00 | 20.00 | 40.00 | [26] |
| VC/Nb/F1.0GDE | -2.5 V | 907.61 | 14.85 | 63.55 | This work |
| Nb1GDE | -1.9 V | 365.00 | 21.00 | 33.00 | [26] |
| VC/Nb/F1.0GDE | -1.9 V | 605.50 | 15.37 | 49.10 | This work |
| Nb1GDE | -1.3 V | 241.00 | 13.00 | 42.00 | [26] |
| VC/Nb/F1.0GDE | -1.3 V | 477.81 | 8.95 | 63.72 | This work |
| Nb1GDE | -0.7 V | 82.00 | 9.50 | 43.00 | [26] |
| VC/Nb/F1.0GDE | -0.7 V | 136.57 | 8.17 | 48.98 | This work |

## 4. Conclusions

In this study, we synthesized F-doped $Nb_2O_5$ in different molar amounts (0, 0.5, 1.0, and 2.0%). XRD analyses showed that the synthesized materials are in agreement with the orthorhombic phase of $Nb_2O_5$ and that no secondary phase was detected, indicating an effective doping of F in the crystal structure of $Nb_2O_5$. Furthermore, XRD analyses showed that the F-doped materials had their crystal lattice expanded, as a consequence of the appearance of $Nb^{4+}$ for charge compensation. SEM and TEM analyses showed that the synthesized materials doped with F maintained the same morphologies as pure $Nb_2O_5$, being the morphology defined as nanorods, as well as the presence of $Nb_2O_5$ based on the identification of the 001 plane of $Nb_2O_5$. The presence of F was also confirmed by EDS elemental mapping analysis, demonstrating that F is homogeneously distributed. The synthesized and characterized materials were then impregnated in Vulcan XC72 carbon and characterized by contact angle and XPS, in addition to being tested for ORR. All $Nb_2O_5$-based electrocatalysts doped with F presented a greater amount of oxygenated species, in particular the $Nb_2O_5$ electrocatalyst doped with 1% presented the greatest amount of oxygenated species, as well as the lowest contact angle value, indicating that it is the most hydrophilic among those synthesized. With the XPS analysis, it was also possible to confirm the presence of $Nb^{4+}$ predicted in the XRD analysis. In the ORR tests, all F-doped $Nb_2O_5$ electrocatalysts exhibited superior performance for the 2-electron mechanism compared to Vulcan XC72 carbon and the carbon electrocatalyst containing pure $Nb_2O_5$, highlighting that F doping was beneficial for the selectivity of the 2-electron mechanism for ORR. Among the electrocatalysts studied, the electrocatalyst containing 1.0% F (VC/Nb/F 1.0%) was the one that presented the best results for ORR, being the one chosen for the preparation of the GDE. The experiments of electrogeneration of $H_2O_2$ using this produced GDE showed that doping with 1.0% F can increase the electrogeneration of $H_2O_2$ by up to 98% compared to the undoped $Nb_2O_5$ catalyst, at the applied potential of -1.3 V. Furthermore, the energy consumption and current efficiency were lower and higher, respectively, when compared to the undoped $Nb_2O_5$ electrocatalyst at all applied potentials, evidencing the greater cost-benefit of this catalyst. The efficiency of the catalyst based on Vulcan carbon XC72 modified with $Nb_2O_5$ doped with 1.0% F is related to the insertion of F in the crystal lattice of $Nb_2O_5$ as a dopant, which caused an increase in the crystal lattice of $Nb_2O_5$ by the presence of $Nb^{4+}$ and local distortions, increasing the degree of disorder of this electrocatalyst, in addition to having increased the number of free electrons to act in the ORR. The F-doped electrocatalysts also presented a greater number of oxygenated species and a more hydrophilic character, which may have facilitated electron transfer, transport and $O_2$ adsorption, improving the electrogeneration of $H_2O_2$ compared to the undoped electrocatalyst.

**CRediT authorship contribution statement**

**Aline B. Trench,** Conceptualization**,** Investigation, Validation, Data curation, Writing – original draft, review & editing**, João Paulo C. Moura, Caio Machado Fernandes:** Investigation, Validation, Data curation, Writing – original draft, review & editing. **Mauro C. Santos:** Conceptualization, Writing – review & editing, Supervision.

**Acknowledgments**

The authors are grateful to the following Brazilian research financing institutions: São Paulo Research Foundation (FAPESP), Brazil, process number, #2021/14394-7, #2021/05364-7, and # 2022/10484-4 for the financial assistance provided in support of this work.